\documentclass[12pt]{article}
\usepackage{amsmath}
\usepackage{cite}
\usepackage{epsfig}
\usepackage{color}

\newdimen\nude\newbox\chek
\def\slash#1{\setbox\chek=\hbox{$#1$}\nude=\wd\chek#1{\kern-\nude/}}

\begin{document}

\def\la{\mathrel{\mathpalette\fun <}}
\def\ga{\mathrel{\mathpalette\fun >}}
\def\fun#1#2{\lower3.6pt\vbox{\baselineskip0pt\lineskip.9pt
  \ialign{$\mathsurround=0pt#1\hfil##\hfil$\crcr#2\crcr\sim\crcr}}}

\def\half{{\textstyle \frac12}}
\def\abs#1{\left|#1\right|}
\def\E{\epsilon}
\def\cO#1{{\cal{O}}\left(#1\right)}
\def\Re{\mathop{\rm Re}}
\def\lrang#1{\left\langle#1\right\rangle}
\def\fm{\mathop{\rm fm}}
\def\GeV{\mathop{\rm Ge\!V}}
\def\MeV{\mathop{\rm Me\!V}}
\def\qhot{\hat{q}_{\mbox{\scriptsize hot}}}
\def\qcold{\hat{q}_{\mbox{\scriptsize cold}}}
\def\as{\alpha_s}
\def\oBH{\omega_{\mbox{\scriptsize BH}}}

 \newskip\humongous \humongous=0pt plus 1000pt minus 1000pt
 \def\caja{\mathsurround=0pt} \def\eqalign#1{\,\vcenter{\openup1\jot
 \caja   \ialign{\strut \hfil$\displaystyle{##}$&$
 \displaystyle{{}##}$\hfil\crcr#1\crcr}}\,} \newif\ifdtup
 \def\panorama{\global\dtuptrue \openup1\jot \caja
 \everycr{\noalign{\ifdtup \global\dtupfalse     \vskip-\lineskiplimit
 \vskip\normallineskiplimit      \else \penalty\interdisplaylinepenalty \fi}}}
 \def\eqalignno#1{\panorama \tabskip=\humongous
 \halignto\displaywidth{\hfil$\displaystyle{##}$
 \tabskip=0pt&$\displaystyle{{}##}$\hfil
 \tabskip=\humongous&\llap{$##$}\tabskp=0pt     \crcr#1\crcr}}

\newcounter{hran}
\renewcommand{\thehran}{\arabic{hran}}

\def\bmini{\setcounter{hran}{\value{equation}}
\refstepcounter{hran}\setcounter{equation}{0}
\renewcommand{\theequation}{\thehran\alph{equation}}\begin{eqnarray}}

\def\bminiG#1{\setcounter{hran}{\value{equation}}
\refstepcounter{hran}\setcounter{equation}{-1}
\renewcommand{\theequation}{\thehran\alph{equation}}
\refstepcounter{equation}\label{#1}\begin{eqnarray}}

%
%
\def\emini{\end{eqnarray}\relax\setcounter{equation}{\value{hran}}\renewcommand{\theequation}{\arabic{equation}}}

\begin{flushright}
BI-TP 2001/02 \\
LPT-Orsay-01/57
\end{flushright}
\vskip1.5cm

\begin{center}
{\Large\bf 
   Quenching of hadron spectra in media}
\end{center}

\begin{center}
{\large  R.~Baier}\\
{\em Fakult\"{a}t f\"{u}r Physik, Universit\"{a}t Bielefeld}, \\
{\em D-33501 Bielefeld, Germany}\\[2mm]
{\large  Yu.L.~Dokshitzer}\footnote{on leave from PNPI, 
St. Petersburg, Russia}\\ 
{\em LPTHE, Universit{\'e} Pierre et Marie Curie (Paris VI)},\\
{\em  4, Place Jussieu, 75252 Paris, France}
\\[2mm] 
{\large  A.H.\ Mueller}\footnote{on sabbatical leave from Columbia
University,  New York, NY, USA}$^,$\!\!
\footnote{supported in part by the US Department of Energy} 
\\[2mm] 
{\large  D. Schiff }\\
{\em LPT, Universit\'e Paris-Sud, B\^atiment 211}, \\
{\em F-91405 Orsay, France}
\end{center}

\begin{abstract}
We determine how the yield of large transverse momentum hadrons is
modified due to induced gluon radiation off a hard parton traversing a
QCD medium.  The quenching factor is formally a collinear- and
infrared-safe quantity and can be treated perturbatively.  In spite of
that, in the $p_\perp$ region of practical interest, its value turns
out to be extremely sensitive to large distances and can be used to
unravel the properties of dense quark-gluon final states produced in
heavy ion collisions.  We also find that the standard modelling of
quenching by shifting $p_\perp$ in the hard parton cross section by
the mean energy loss is inadequate.
\end{abstract}

\section{Introduction}
The so-called jet quenching
\cite{ref1,ref2,ref3,ref4a,ref4b,ref_descr} is considered an important
signal of the production of a new state of dense matter (quark-gluon
plasma) in ultrarelativistic heavy ion collisions.  This is understood
as the suppression of the yield of large transverse momentum jets or
particles with respect to proton-proton collisions.

In this paper we concentrate on the quenching effect in inclusive
particle spectra, due to the energy loss by medium induced gluon
radiation
\cite{ref5,ref6,ref7,ref8,ref9,ref10,ref11,ref12,ref13,ref14,ref15,ref16,ref16a,ref17,ref18,ref19,ref19a,ref22}.

Inclusive production of particles with large $p_\perp$, in, say,
proton-proton collisions can be parametrized as a power
\begin{equation}
\label{eq:sigvac}
\frac{d\sigma^{{\rm vacuum}}(p_\perp)}{dp_\perp^2} \propto \frac1{p_\perp^n}\,,
\qquad 
 n=n(p_\perp)
   \>\equiv\> - \frac{d}{d\ln p_\perp} \> 
   \ln \frac{d\sigma^{{\rm vacuum}}(p_\perp)}{dp_\perp^2}
\end{equation}
with $n$ an effective exponent which slowly decreases with increase of
$p_\perp$.  In reality, for moderately large $p_\perp$ the exponent in
\eqref{eq:sigvac} is much larger than the asymptotic value $n=4$
corresponding to the limit $p_\perp\to\infty$
($s/p_\perp^2=\mbox{const}$).

The effects that add to the value of $n$ are:
\begin{enumerate}
\item 
 $x$-dependence of the parton distributions which decrease with
 increase of parton energies, $x_1,x_2\propto \sqrt{p_\perp^2/s}$, 
\item 
 the bias effect due to vetoing accompanying gluon radiation off the
 primary large-$p_t$ parton that produces the triggered particle with
 $p_\perp\le p_t$, 
\item 
 running of the coupling $\alpha_s^2(p^2_\perp)$ in the parton-parton
 scattering cross section.
\end{enumerate}
As a result, in practice $n$ is seen to be as large as 10.  In what
follows we shall treat $n$ as a large numerical parameter neglecting
relative corrections of the order of $1/n$.  (Such an approximation,
though unnecessary, allows us to derive a simple analytic expression
for jet quenching due to medium effects.)

In the presence of a medium the inclusive spectrum
\eqref{eq:sigvac} changes. 
Multiple interactions of partons with the medium lead to two competing
effects.  One the one hand, multiple scattering in the initial (as
well as in the final) state partially transforms longitudinal parton
motion into transverse one, thus enhancing the yield of
large-$p_\perp$ particles (the so-called Cronin effect~\cite{Cronin}).
On the other hand, medium induced gluon radiation accompanying
multiple scattering causes parton energy loss and therefore suppresses
the particle yield.

At large transverse momenta the second effect takes over.

To find the inclusive particle spectrum, one has to convolute the
production cross section of the parton with energy $p_\perp+\E$ with
the distribution $D(\E)$ in the parton energy loss $\E$ in the final
state:\footnote{For the sake of simplicity we consider particle
production at $90^{\mbox{\scriptsize o}}$ and equate the transverse
momentum of the particle $p_\perp$ with its energy.}
\begin{equation} 
\label{eq:convol}
 \frac{d\sigma^{{\rm medium}}(p_\perp)}{dp^2_\perp} =
\int d\E \, D(\E) \, \frac{d\sigma^{{\rm vacuum}} (p_\perp + \E)}{dp^2_\perp}. 
\end{equation}
Since the effects of the standard jet fragmentation which reduce the
energy of the leading particle in the jet are already included in the
vacuum cross section, the distribution $D(\E)$ here describes
specifically the {\em additional}\/ energy loss due to medium induced
gluon radiation in the final state.

The quenching effect is customarily modelled by the substitution
\begin{equation} \label{eq:DEmodel}
 \frac{d\sigma^{{\rm medium}}(p_\perp)}{dp^2_\perp} =
 \frac{d\sigma^{{\rm vacuum}} (p_\perp + S)}{dp^2_\perp}. 
\end{equation}
The {\em shift}\/ parameter $S$ in \eqref{eq:DEmodel} is usually taken
either proportional to the size of the medium,
\bminiG{models}
\label{eq:naive}
   S \>=\> \mbox{const}\cdot L \,,
\end{eqnarray}
or equal to the {\em mean}\/ medium induced energy loss~\cite{ref9}
\begin{eqnarray} 
\label{eq:meanlossmod}
   S\>=\> \Delta E \equiv \int d\E\, \E\, D(\E)\> \propto\>
   \as \,L^2\,.
\emini
The former ansatz has no theoretical justification while the latter
emerges as a result of the Taylor expansion of \eqref{eq:convol} based
on the $\E\ll p_\perp$ approximation:
\begin{equation}
\eqalign{
\int d\E \, D(\E) \cdot \frac{d\sigma(p_\perp+\E)}{dp^2_\perp}
& = \int d\E \, D(\E) \cdot \frac{d\sigma(p_\perp)}{dp^2_\perp} + \int
d\E \,\E\, D(\E)
\cdot \frac{d}{d p_\perp}\frac{d\sigma(p_\perp)}{dp^2_\perp} + \ldots \cr
& \simeq \frac{d\sigma}{dp^2_\perp} +
\Delta E\cdot \frac{d}{d p_\perp}\left(\frac{d\sigma}{dp^2_\perp}\right) 
\simeq \frac{d\sigma (p_\perp+\Delta E)}{dp^2_\perp} .
}
\end{equation}
Such an approximation misses, however, one essential point, namely
that the vacuum distribution is a sharply falling function of
$p_\perp$. This causes a strong bias which leads to an additional
suppression of real gluon radiation. As a result, the {\em typical}\/
energy carried by accompanying gluons turns out to be much smaller
than the {\em mean}\/ \eqref{eq:meanlossmod}.

In this paper we study an interplay between the energy loss and the
cross section fall-off and show that it leads, in the region of
transverse momenta of practical interest, to the $p_\perp$ dependent
expression for the shift
\[
   S(p_\perp) \>\propto\>  \sqrt{ p_\perp}\,.
\]

\section{Medium induced energy loss}

\subsection{Transport coefficient}
The two effects --- parton transverse momentum broadening and medium
induced radiation are closely related and are determined by the
so-called ``transport coefficient'' $\hat{q}$ which characterizes the
``scattering power'' of the medium~\cite{ref9}:
\begin{equation}\label{eq:qdef}
  {\hat{q}}^{(R)} = \rho \int dq^2\, q^2 \, \frac{d\sigma^{(R)}}{dq^2} \,.
\end{equation}
Here $\rho$ is the density of scattering centres, and $d\sigma^{(R)}$
is the single scattering cross section for a projectile parton in the
colour representation $R$, with $C_R$ the corresponding colour factor
($C_F=(N_c^2-1)/2N_c= 4/3$, $C_A=N_c=3$ for quark and gluon,
respectively).

The dimensionless ratio $\hat{q}/\rho$ characterizes the ``opacity''
of the medium for an energetic gluon. It includes the region of small
momentum transfers where the perturbative treatment is hardly
applicable, and should be regarded as (the only) unknown
medium-dependent parameter of the problem.

In ``cold'' nuclear matter $\hat{q}$ can be calculated perturbatively
and related to the gluon density $[xG(x,Q^2)]$ of the nucleus at a
low momentum scale $Q^2\simeq \hat{q}\, L$ and small but not too small
$x$, where the gluon density has little dependence on~$x$~\cite{ref9}:
\begin{equation}\label{eq:qcold}
  {\hat{q}}^{(R)} 
  \simeq \rho\, \frac{4\pi^2
  \as C_R}{N_c^2-1} \left[\, xG(x,{\hat{q}}^{(R)} L)\,\right].
\end{equation}
Hereafter we shall label $\hat{q}$ the {\em gluon}\/ transport
coefficient. Taking $\rho=0.16~\fm^{-3}$, $\as=0.5$ and $xG(x)=1$ in
\eqref{eq:qcold} results in
\begin{equation}\label{eq:qcoldest}
 \qcold \>\simeq\> 0.009\, {\GeV}^3 \>\simeq\> 0.045\,  \frac{\GeV^2}{\fm}. 
\end{equation}
$\hat{q}$ enters, in particular, as the proportionality factor between
an accumulated parton transverse momentum squared and the size of the
medium traversed, $\kappa^2\,\propto\, \hat{q} L$.  The quark
transport coefficients extracted from experimentally measured
transverse momentum nuclear broadening of Drell-Yan lepton
pairs~\cite{DY} agrees with the theoretical
estimate~\eqref{eq:qcoldest}.

In the case of heavy ion collisions, the scattered hard parton
traverses a medium that is expected to have an energy density much
higher than that of nuclear matter, and the corresponding transport
coefficient $\qhot$ can be much larger. If hot matter is formed in
the final state, a perturbative estimate for the QGP with
$T=250\,\MeV$ gives~\cite{ref9}
\begin{equation}\label{eq:hotest}
   \qhot \>\simeq\> 0.2\, {\GeV}^3 \>\simeq\> 1\,  \frac{\GeV^2}{\fm}. 
\end{equation}

\subsection{Induced gluon radiation}

To find the distribution $D(\E)$ in the parton energy loss $\E$ in the
final state we need to recall the basic properties of gluon radiation 
caused by multiple parton scattering in the medium.

Introducing the characteristic gluon frequency 
\begin{equation}  \label{eq:omega_1}
  \omega_c = \frac{\hat{q}}{2}\, L^2 \,,
\end{equation}
with $L$ the length of the medium, 
the inclusive energy spectrum of medium induced soft gluon radiation 
($\omega\ll p_\perp$) reads~\cite{ref10,ref11} 
\begin{equation}
\label{eq:dIdo}
 \frac{dI(\omega)}{d\omega} =\frac{\alpha}{\omega} \ln
 \abs{\cos\left[\,(1+i)\sqrt{\frac{\omega_c}{2\omega}}\,\right]} 
= \frac{\alpha}{2\omega} \,
\ln \left[\,\cosh^2\sqrt{\frac{\omega_c}{2\omega}} -
\sin^2\sqrt{\frac{\omega_c}{2\omega}} \,\right]; \qquad
\alpha\equiv \frac{2\alpha_s C_R}{\pi}. 
\end{equation}
This distribution peaks at small gluon energies, 
\begin{equation}
\label{eq:dI_small}
 \omega \frac{dI(\omega)}{d\omega} 
= \alpha \left\{ \sqrt\frac{\omega_c}{2\omega}
- \ln 2 \right\}
\cdot\left[\,1 +
\cO{\exp\left\{-\sqrt{\frac{2\omega_c}{\omega}}\right\}}\right], 
\qquad \omega < \omega_c\, ,
\end{equation}
while for energies above the characteristic scale it is small and
falling fast with $\omega$:
\begin{equation}
\label{eq:dI_large}
\omega  \frac{dI(\omega)}{d\omega}  \simeq \frac{\alpha}{12}
\left(\frac{\omega_c}{\omega}\right)^2 , \qquad \omega > \omega_c\,.
\end{equation}
The multiplicity of gluons with energies larger than a given $\omega$
is given by the integal of the inclusive gluon spectrum
\eqref{eq:dIdo}:
\begin{equation}\label{eq:mult}
\begin{split}
  N\left( {\omega}\right) \equiv \int_\omega^\infty d\omega'\, \frac{dI(\omega')}{d\omega'}
&= \alpha\int_0^{\sqrt{\omega_c/2\omega}} \frac{dz}{z} 
 \ln \left(\cosh^2z - \sin^2z\right).
\end{split}
\end{equation}
Evaluating this integral for $x=\omega/\omega_c\ll 1$ we have
(cf.\ \eqref{eq:dI_small})
\begin{equation}
\label{eq:Nexpr}
   N(\omega) \simeq \alpha\left[\,  \sqrt\frac{2}{x} +\ln2\ln x
- 1.44136 \>+\> \cO{\exp(-\sqrt{2/x})} \right],
\end{equation}
with the constant term found by numerical integration of the exact
spectrum \eqref{eq:dIdo}.

\subsection{Energy loss distribution}

Since the vacuum spectrum \eqref{eq:sigvac} is falling fast with
$p_\perp$, the bias effect forces the distribution $D(\E)$ to the
smallest energy losses possible, $\E/p_\perp\ll 1$. In these
circumstances the ``final'' parton is the one that had been produced
in the hard interaction: the quark-gluon transition is additionally
suppressed as $\E/p_\perp$ and can be neglected. Bearing this in
mind, we will treat $D(\E)$ as the distribution in energy that the
hard parton (a quark or a gluon) loses to medium induced gluon
bremsstrahlung.

The spectrum of the leading particle can be characterised by the
probability $D(\E)$ that the radiated gluons carry altogether a given
energy $\E$. The corresponding expression based on independent
emission of soft primary gluons reads
\begin{equation}  \label{eq:ind}
D(\E) = \sum^\infty_{n=0} \, \frac{1}{n!} \,
\left[ \prod^n_{i=1} \, \int \, d\omega_i \, \frac{dI(\omega_i)}{d \omega} 
\right] \delta \left(\E - \sum_{i=1}^n  \omega_i\right)
\cdot \exp \left[ - \int d\omega \frac{dI}{d\omega} \right], 
\end{equation} 
where the last factor accounts for virtual effects. 
In a standard way, the energy constraint can be factorized using
the Mellin representation,
\begin{equation}
\label{eq:deltaMell}
 \delta\left(\E - \sum_{i=1}^n \omega_i\right) = \int_C
 \frac{d\nu}{2\pi i} \> e^{\nu {\E}}\cdot \prod_{i=1}^n  e^{-\nu{\omega_i}},
\end{equation}
after which multiple gluon radiation exponentiates, and the answer can
be written as
\bminiG{eq:Mells}
\label{eq:DMelldef}
D(\E) &=& \int_C \, \frac{d\nu}{2\pi i} \>
\tilde{D}(\nu)\,e^{\nu {\E}}, \\
\label{eq:DMell}
 \tilde{D}(\nu) &=& \exp \left[ - \int^\infty_0 d\omega \, 
\frac{dI(\omega)}{d\omega} \left( 1 - e^{-\nu\omega}\right)
\right]. 
\emini
The contour $C$ in \eqref{eq:deltaMell} and \eqref{eq:DMelldef} runs
parallel to the imaginary axis in the complex $\nu$-plane,
$\Re\nu=\mbox{const}$.  An equivalent expression can be written in
terms of the integrated gluon multiplicity \eqref{eq:mult}.
Integrating \eqref{eq:DMell} by parts, we obtain an elegant formula
\bminiG{eq:Delegant}
\label{eq:elegant1}
 \tilde{D}(\nu)  &\>=\>& 
\exp \left[ -\nu\int^\infty_0 d\omega\,e^{-\nu\omega} \, 
N\left(\omega\right) \right] \\
\label{eq:elegant2}
&\>=\>&  \exp \left[ -\int^\infty_0 dz\,e^{-z} \, 
N\left(\frac{z}{\nu}\right) \right].
\emini
As we shall see shortly, the value of the dimensional 
variable $\nu\omega_c$
is linked with the characteristic parameter of the problem
$n\omega_c/p_\perp$ which is typically much larger than unity.
Therefore, in the essential integration region in
\eqref{eq:DMell} we have $\omega\sim 1/\nu\sim p_\perp/n \ll\omega_c$,
and the approximate expression \eqref{eq:Nexpr} can be used. This
gives
\begin{equation} 
\label{tDappr}
 \tilde{D}(\nu) 
\>\simeq\> \exp \left\{ -\alpha\left(\sqrt{2\pi \nu\omega_c}  
- \ln2 \ln(\nu\omega_c) - 1.84146 \right)  \right\} , 
\end{equation}
which expression allows one to evaluate the energy loss distribution
$D(\E)$ for $\E\ll \omega_c$ analytically in terms of the
hypergeometric function.

For the purpose of illustration we present here only a rough estimate
based on the leading small-energy behaviour of the gluon distribution
$\omega dI/d\omega \propto \as\sqrt{\omega_c/\omega}$ which translates
into
\begin{equation} 
\label{tDest}
 \tilde{D}(\nu) 
\>\simeq\>  \exp \left\{ -\alpha \sqrt{2\pi \nu\omega_c} \right\}.  
\end{equation}
The Laplace integral \eqref{eq:DMelldef} then becomes Gaussian and
yields
\begin{equation}
\label{eq:Dappr}
\E D(\E) \>\simeq\> {\alpha}\sqrt{\frac{\omega_c}{2\E}} \>
\exp\left\{-\frac{\pi\alpha^2\omega_c}{2\E}\right\}.
\end{equation}
This approximation is fine for illustrative purposes, in particular
because the distribution \eqref{eq:Dappr} is properly normalized to
unity due to $\tilde{D}(\nu=0)=1$ in \eqref{tDest} .
As expected, in the first order in $\as$ the energy loss spectrum
coincides with the probability of emission of one gluon with the
energy $\omega=\E$, see \eqref{eq:dI_small}.  The one-gluon
approximation spectacularly fails, however, at small $\E$. If the
energy loss is taken as small as $\E\sim \alpha^2\omega_c$, the
distribution reaches its maximum at
\begin{equation}
\label{eq:Dmax}
 {\E} \>=\> \pi\alpha^2\,{\omega_c}\> \ll\> \omega_c
\end{equation}
and becomes exponentially small at smaller energies, due to form
factor suppression.

\section{Quenching}

We introduce the medium dependent {\em quenching factor}\/ $Q$ to
represent the inclusive particle spectrum as
\begin{equation}
\frac{d\sigma^{{\rm medium}}(p_\perp)}{dp^2_\perp} =
 \frac{d\sigma^{{\rm vacuum}} (p_\perp)}{dp^2_\perp}\cdot Q(p_\perp) 
\end{equation}
with
\begin{equation}\label{eq:Qdef}
Q(p_\perp) \>=\> \int d\E\, D(\E)\cdot \left(\frac{d\sigma^{{\rm vacuum}}
(p_\perp+\E)/dp_\perp^2}{d\sigma^{{\rm vacuum}}
(p_\perp)/dp_\perp^2}\right). 
\end{equation}
Using the fact that the effective exponent $n$ in the ratio
\[
R\>=\> \frac{d\sigma^{{\rm
vacuum}}(p_\perp+\E)/dp_\perp^2}{d\sigma^{{\rm vacuum}}
(p_\perp)/dp_\perp^2} \>\simeq\>
\left(\frac{p_\perp}{p_\perp+\E}\right)^n
\]
entering the convolution \eqref{eq:Qdef} is numerically large, we can
replace $R$ by the exponential form
\begin{equation}\label{eq:n_approx}
R  \>=\> \exp\left(-\frac{n\E}{p_\perp}\right)\cdot
\left[\,1+\cO{\frac{\ln^2 R}{2\, n}}\right] .
\end{equation}
The accuracy of such a substitution is rather good all over the region
of practical interest where the quenching suppression is moderately
large.

The exponential approximation \eqref{eq:n_approx}
results in a simple expression for the quenching, as the suppression
factor equals a given Mellin moment in $\E$ of the spectrum $D(\E)$
(see \eqref{eq:DMelldef} and \eqref{eq:Delegant}):
\begin{equation} 
\label{eq:Fdef}
Q(p_\perp) \simeq \int_0^\infty d\E \, D(\E)\,
\exp\left\{-\frac{n}{p_\perp}\cdot\E\right\}
\>=\> \tilde{D}\left(\frac{n}{p_\perp}\right) \>=\>\exp\left\{
-\int_0^\infty dz\,e^{-z}\> N\left(\frac{p_\perp}{n}z\right)\right\}.
\end{equation}
Representing the quenching factor as
\bminiG{eq:supprfac}
 Q(p_\perp) &=& \exp\left\{-\frac{n}{p_\perp}\cdot
S(p_\perp)\right\}, \\
\label{eq:Sint}
 S(p_\perp) &\equiv& -\frac{p_\perp}{n}\,\ln
\tilde{D}\left(\frac{n}{p_\perp}\right)
= \int_0^\infty d\omega\, N(\omega) \,
\exp\left\{- \frac{n\,\omega}{p_\perp} \right\}, 
\emini
and using again the large-$n$ approximation, we can cast the quenching
effect as a {\em shift}\/ of the vacuum spectrum:
\begin{equation} 
\label{eq:shiftdef}
\eqalign{
\frac{d\sigma^{{\rm medium}}(p_\perp)}{dp^2_\perp}  &\simeq
\frac{d\sigma^{{\rm vacuum}}(p_\perp+S(p_\perp))}{dp^2_\perp}\,.
}
\end{equation} 
It should be noted that the shift approximation
\eqref{eq:shiftdef} {\em underestimates}\/ suppression 
if the quenching factor $Q$ becomes very small.

The general expression \eqref{eq:Fdef} actually solves the quenching
problem, given the gluon emission spectrum $dI/d\omega$.
To verify the validity of the approach and the approximations made, we
need, however, to estimate the characteristic gluon energies that
determine the answer. To this end we proceed with a qualitative
analysis of QCD quenching.

Using the small-energy approximation \eqref{eq:Nexpr} in
\eqref{eq:Fdef} we derive
\begin{equation}
\label{eq:simple}
   Q(p_\perp) \>=\> \exp\left\{ -\sqrt{\pi}\,
   N\left(\frac{p_\perp}{n}\right) \right\}
\>\simeq\> \exp\left\{ - N\left(\frac{p_\perp}{\pi\, n}\right) \right\}. 
\end{equation}
This expression has a simple physical interpretation.  It is a typical
exponential form factor suppression whose exponent equals the
probability of gluon radiation in the {\em forbidden}\/ kinematical
region or, in other words, the mean multiplicity of virtual gluons.
We see that the energies of {\em real}\/ gluons are cut from above as
\begin{equation}
\label{eq:realcut}
  \omega \><\> \frac{\omega_1}{\pi} \,, \qquad \omega_1= \frac{p_\perp}{ n}\,, 
\end{equation}
which (modulo a constant\footnote{dimensional, some would add,} factor
$\pi$) equals a small ($1/n$) fraction of the energy of the registered
particle, due to the bias effect.
The characteristic energy parameter of the quenching problem can be
directly expressed in terms of the hard vacuum cross section by the
general relation 
\begin{equation}
\label{eq:omega1gen}
  \omega_1(p_\perp) \>\equiv\> -\left[\> \frac{d}{dp_\perp}\,  
\ln \frac{d\sigma^{{\rm vacuum}}}{dp_\perp^2} \>\right]^{-1},
\end{equation}
which does not rely on the power approximation \eqref{eq:sigvac} with
$n=$const.

For the ``shift'' function \eqref{eq:supprfac} in the approximation
\eqref{eq:Nexpr} we obtain
\begin{equation}
\label{eq:shiftans}
  S(p_\perp) \>\simeq\> \sqrt{\frac{2\pi\,\alpha^2\, \omega_c p_\perp}{n}}\,.
\end{equation}
This result namely, the $p_\perp$ dependent shift, is very different
from the models \eqref{models} that were discussed in the
literature and are being used to predict/describe
quenching~\cite{ref1,ref2,ref3,ref4a,ref4b,ref_descr}.
To understand the origin of \eqref{eq:shiftans} in the rest of this
section we consider and compare the mean energy loss with a typical
energy loss that characterizes quenching.

\paragraph{Mean energy loss.}

The {\em mean}\/ medium induced energy loss of a parton with a
given energy $E$ is determined by the integral
\begin{equation}
\Delta E\equiv 
\int_0^\infty d\E\, \E\, D(\E) \>=\> -\left. \frac{d}{d\nu}
\tilde{D}(\nu)\right|_{\nu=0} \>=\> \int_0^\infty d\omega\,\omega\,
\frac{dI(\omega)}{d\omega} \>=\> \int_0^\infty d\omega\, N(\omega),
\end{equation}
where we have used \eqref{eq:Mells}.
Given $\omega dI/d\omega\propto N(\omega)\propto
\omega^{-1/2}$, this integral is determined by the largest available
gluon energies, $\omega\la\omega_{\max}$, resulting
in~\cite{ref9}
\begin{equation}
\label{eq:avloss}
 \Delta E \propto \alpha_s \sqrt{\omega_c\,\omega_{\max}}\,, \qquad 
 \omega_{\max} \>=\> \min\left\{\> \omega_c\,,\> E\>\right\}.
\end{equation}

The commonly used identification of the shift with the mean energy
loss \eqref{eq:meanlossmod} is valid in the region of large particle
energies $p_\perp\ga n\omega_c$, corresponding to the quenching
parameter $\nu\omega_c\la 1$.
Indeed, since the integrated gluon multiplicity $N(\omega)$ vanishes
fast for $\omega>\omega_c$, for $\nu<1/\omega_c$ we can omit the
exponential factor in the integrand of \eqref{eq:elegant1} to derive
\begin{equation} 
 S(p_\perp)\simeq \int_0^\infty d\omega\, N(\omega) 
\equiv \Delta E\,.
\end{equation}
In this region, however, the quenching itself is vanishingly weak
since $\Delta E\propto\as\omega_c$ and
\[
  -\ln Q(p_\perp) = \frac{n}{p_\perp}\cdot \Delta E \>\>\propto\>\> 
  \as\cdot \frac{n\omega_c}{p_\perp} < \as\,.  
\]
Gluons responsible for the mean energy loss are rare, $\cO{\as}$.
Strictly speaking, rare fluctuations with energetic gluons ($\omega\ga
\omega_c$) in $dI/d\omega$ do contribute to quenching via the virtual
suppression factor in \eqref{eq:DMelldef}:
$$ 
\delta \ln Q\>=\>  -\int_{\omega_{c}}^\infty d\omega\,
\frac{dI(\omega)}{d\omega} .
$$
This contribution, however, is bound to be small (negligible) as
long as the mean {\em multiplicity}\/ (not mean {\em energy}\/!) 
of such gluons is of the order $\as$.

This argument applies both to the ``canonical'' $\Delta
E$~\cite{ref10,ref11} and to the additional contribution to $\Delta E$
originating from emission of energetic gluons when there is only a
{\em single}\/ scattering in the medium, found in a series of recent
papers initiated by Gyulassy, L\'evai and
Vitev~\cite{ref18,ref19,ref19a,ref17}.
The GLV energy spectrum has an enhanced high-energy tail, 
$$
 \frac{\omega\, dI^{(\mbox{\scriptsize GLV})}(\omega)}{d\omega} \propto 
 \alpha\,\frac{\omega_c}{\omega}\,, \quad \omega>\omega_c \qquad
\left[\> \mbox{ cf.}\qquad 
\frac{\omega\, dI^{(\mbox{\scriptsize BDMPS})}(\omega)}{d\omega} \propto 
 \alpha\left(\frac{\omega_c}{\omega}\right)^2 \>\right], 
$$
thus inducing a potentially large contribution to $\Delta E$ from the
region $\omega\gg\omega_c$.  Such fluctuations, however, do not affect
quenching since
$$
 \int_{\omega_c}^\infty d\omega\, 
 \frac{dI^{(\mbox{\scriptsize GLV})}(\omega)}{d\omega} \>=\>\cO{\as}.
$$

\paragraph{Typical energy loss.}

To have a significant quenching, $-\ln Q(p_\perp)=\cO{1}$, 
we have to have $N(p_\perp/n)\sim 1$, according to
\eqref{eq:Fdef}, which translates into $p_\perp \la \alpha^2
n\omega_c \ll n\omega_c$.   

The value of the convolution integral \eqref{eq:convol} that determines
quenching results in an interplay of the steep fall-off of the parton
cross section with $\E$ and the form factor suppression of small
losses. To estimate the characteristic energy loss for a given
$p_\perp$ we invoke the approximate expression \eqref{eq:Dappr} for
the
distribution $D(\E)$ to write
\bminiG{2Qs}
\label{eq:Fest}
Q(p_\perp) &\simeq& \frac{\alpha}{\sqrt{2}} \int_0^\infty
\frac{dx}{x^{\frac32}}
\exp\left\{-\frac{\pi\alpha^2}{2\,x}
-\frac{n\omega_c}{p_\perp}x \right\} 
= \exp\left\{ -2\sqrt{\frac{\pi\alpha^2 n\omega_c}{2\,p_\perp}}\right\}, 
\quad
x\equiv\frac{\E}{\omega_c}. \qquad { }   
\end{eqnarray}
In the region where the quenching is strong, $Q\ll 1$, the steepest
descent evaluation applies, provided the exponent on the r.h.s.\ of
\eqref{eq:Fest} is large.

With $p_\perp$ increasing, quenching becomes week, $1-Q(p_\perp)\ll
1$. In this kinematical region we use the fact that $D(\E)$ is
normalized to unity to write an equivalent representation
\begin{eqnarray} 
\label{eq:Festeq}
1-Q(p_\perp) &\simeq& \sqrt{\frac{\alpha^2n\omega_c}{2\,p_\perp}}
 \int_0^\infty \frac{dy}{y}\,
\frac{1- e^{-y}}{\sqrt{y}}\cdot 
\exp\left\{-\frac{\pi\alpha^2n\omega_c}{2p_\perp\,y}\right\}  \,, \quad 
y=\frac{n\,\E}{p_\perp}.
\emini
Now, if the exponent in the last factor is small it can be dropped,
and the integral over $y$ is determined by $y=\cO{1}$. 

Combining the two estimates we arrive at
\bminiG{eq:border}
\label{eq:smallX}
 \lrang{\E} &\simeq& \sqrt{\frac{\pi \alpha^2\omega_c\,p_\perp}{2\, n} }\,
\qquad   \mbox{for} \qquad  {p_\perp} <
\frac\pi2\, \alpha^2\,{n\,\omega_c}\,, \\
\label{eq:largeX}
 \lrang{\E} &\simeq& 
\qquad \frac{p_\perp}{n}\qquad\qquad
\mbox{for}\qquad   
{p_\perp} >  \frac\pi 2\,\alpha^2\,{n\,\omega_c}\,,
\emini
where $\lrang{\E}$ is the typical energy value dominating the
integrals \eqref{2Qs}.
The two expressions \eqref{eq:border} match at the border value.

\paragraph{Typical energy loss and the shift.}

In the first regime \eqref{eq:smallX} gluon radiation is omnipresent.
Invoking \eqref{eq:supprfac} we observe that the characteristic energy
loss $\lrang{\E}$ in \eqref{eq:smallX} equals {\em half}\/ of the
shift function $S(p_\perp)$.  The factor $\half$ may look
anti-intuitive at a first glance. To appreciate it we need to recall
that the substitution
\[
\frac{d\sigma^{{\rm medium}}(p_\perp)}{dp^2_\perp}
\>\Longrightarrow\> 
\frac{d\sigma^{{\rm vacuum}} (p_\perp + \lrang{\E})}{dp^2_\perp}    
\]
accounts only for a part of quenching: the second ingredient of the
convolution \eqref{eq:convol} --- the energy loss distribution 
$D\left(\lrang{\E}\right)$ --- also supplies an (equal) exponential
suppression factor, giving
\begin{equation}
\label{eq:shiftappr}
  S(p_\perp) \> = \> 2\lrang{\E} \>=\> 2\cdot 
  \sqrt{\frac{\pi\alpha^2 \omega_c\, p_\perp}{2\>n}}.
\end{equation}
In the complementary region of larger $p_\perp$, the situation
changes. Here gluon radiation is rare and can be treated as a
correction.  In the one-gluon approximation the loss is equal the
gluon energy, $\lrang{\E}=\omega_1$ in \eqref{eq:largeX}. It has to be
multiplied by the (small) gluon probability (multiplicity) $\Delta
M(\omega_1)$,
\bminiG{eq:M}
\label{eq:Mdef}
  \Delta M(\omega) \>=\>  \mbox{const}\cdot \left[\,
\omega\frac{dI(\omega)}{d\omega}\, \right]
\end{eqnarray}
to obtain the shift \eqref{eq:shiftans},
\begin{eqnarray}
\label{eq:SDM}
  S(p_\perp)\>=\>  \omega_1\cdot \Delta M(\omega_1) \,.
\emini

\paragraph{Characteristic single gluon energy.}

In \eqref{eq:realcut} we have introduced the characteristic (single
gluon) energy $\omega_1$ as a border separating real and virtual
emissions.  Real and virtual gluons {\em softer}\/ than $\omega_1$
cancelled, while the multiplicity of {\em harder}\/ virtual gluons
provided the form factor suppression which resulted in quenching
according to \eqref{eq:Fdef}.  Given an interpretation of $\omega_1$
as the maximal energy of {\em real}\/ gluons, the shift can again be
expressed as in \eqref{eq:SDM} which evaluation is now valid for
arbitrary $p_\perp$, across \eqref{eq:border}.

\section{Approximations}

We start the discussion of the approximations made by noting that
having written the factorised convolution \eqref{eq:convol} we
tacitly supposed that the leading quark turns into the registered
hadron (hadronizes) {\em outside}\/ the medium, which implies
\[
    p_\perp \, R_{\mbox{\scriptsize conf}}^2 \>>\> L\,.
\]
Then, the power approximation for the vacuum spectrum with $n=$const
can be justified.  Our derivation could have been damaged if the
effective exponent $n$ in \eqref{eq:sigvac} changed significantly over
the energy range $p_\perp \div p_\perp+\E$ in the essential region in
$\E$. Substituting the typical energy loss
\eqref{eq:smallX} we observe that it
provides a correction
\[
   p_\perp + \lrang{\E} = p_\perp\left(1 + \frac{\ln Q^{-1}}{2\,n} \right) 
\]
which is actually {\em relatively small}\/ for all values of $Q$ of
practical interest.  This estimate allows us to evaluate $n$ at the
value of the transverse momentum of the registered particle,
$n=n(p_\perp)$, and use the general expression \eqref{eq:omega1gen} as
the definition of the characteristic energy parameter $\omega_1$.

\paragraph{Soft approximation.}

The medium induced gluon distribution \eqref{eq:dIdo} is valid in the
soft gluon approximation, $\omega\ll p_\perp$.  This approximation is
justified by the fact that the main contribution to \eqref{eq:simple}
comes from gluons with energies
\begin{equation}
\label{eq:xtyp}
   \omega \>\sim\>  \omega_1 =\frac{p_\perp}{n} \ll p_\perp\,.  
\end{equation}
Moreover, finite quenching, $\ln Q^{-1}= \cO{1}$, implies $p_\perp\la
\alpha^2\,n\omega_c$ so that 
\begin{equation}
\label{eq:char}
   \omega \>\sim\>  \omega_1 =\frac{p_\perp}{n}  \la \alpha^2\,\omega_c
\ll \omega_c\,,
\end{equation}
so that the approximate expressions \eqref{eq:dI_small},
\eqref{eq:Nexpr} are legitimate.\footnote{Subleading corrections due to hard
gluons will be discussed in the next section.} 

\paragraph{Independent gluon radiation.}

The master equation \eqref{eq:ind} was based on the independent gluon
radiation picture.  To estimate significance of possible interference
effects we need to look at a typical density of gluons in
configuration space.

To this end we recall that the lifetime (formation time) $t$ of a
medium induced gluon with a given energy $\omega$ and transverse
momentum $k_\perp$ follows from the relations
\begin{equation}
 t\sim\frac{\omega}{k^2_\perp}\,, \quad 
 k_\perp^2 \sim \frac{t}{\lambda} \, \mu^2\,,
\end{equation}
where $\mu$ is a typical transverse momentum transfer in a single
scattering and $\lambda$ the gluon mean free path.  This gives
($\mu^2/\lambda \simeq \hat{q}$)
\bminiG{eq:ests}
\label{eq:estt}
 t &\simeq& 
\sqrt{\frac{\omega}{\hat{q}}}\,, \\
\label{eq:estkp}
 k_\perp^2 &\simeq& \sqrt{\omega\,\hat{q}}\,.
\emini
Multiplying \eqref{eq:estt} by the multiplicity density
\eqref{eq:Mdef} and dividing by the size of the medium, we 
estimate the ``gluon occupation number'',
$$ 
 {\rm Occup} \>\sim {t}\cdot {\Delta M}\cdot\frac1{L} \>\sim\>
   \sqrt{\frac{\omega}{\hat{q}}} \>\cdot \>
   \alpha\sqrt{\frac{2\omega_c}{\omega}}  \>\cdot \> \frac1L 
  \> \> = \> \>\alpha \,.
$$
This looks natural: a bare quark produced in a hard interaction
builds up its wave function by emitting gluons, with $\alpha$ being the
gluon density per unit phase space. The r\^ole of a (dense) medium is
to strip the quark off these gluons, so that the formation process
repeats again and again, resulting in induced production of many
gluons.

The smallness of the overlap between gluons\footnote{A similar
argument applies to an interference between {\em medium}- and {\em
vacuum}-produced gluons.}  allows us to neglect possible interference
effects in multiple gluon radiation as potentially contributing at the
level of $\cO{\as}$, while the main effect of the resummed medium
induced radiation in \eqref{eq:ind} is $\cO{1}$.  This justifies,
aposteriori, the Poisson approximation \eqref{eq:ind}.

\paragraph{Scale of the coupling.}

According to \eqref{eq:estkp} the characteristic scale of the QCD
coupling $\as(k_\perp)$ in the gluon emission spectrum
\eqref{eq:dIdo} is rather small. For typical gluon energies
\eqref{eq:realcut} we have ($\hat{q}=0.2\,\GeV^3$)
\[
 k_{1\perp} \sim \left(\frac{\hat{q}}{n}\,p_\perp\right)^{1/4}  
 \>\simeq\> 700\,\MeV\cdot
\left(\frac{p_\perp}{n\cdot\GeV}\,\right)^{1/4}  
\>=\cO{ 1\,\GeV}\,, 
\] 
which value is rather low and marginally increases with $p_\perp$
reaching $k_{1\perp} \sim 1.5\,\GeV$ for $p_\perp\sim 100\,\GeV$
($n\simeq 6$).
In what follows we take a fixed value for the coupling 
\[
    \as \>=\> 0.5\,, \quad \alpha=\frac{2\,\as\,C_F}{\pi} \simeq 0.42\,,
\]
motivated by the studies of the effective QCD interaction strength in
the infrared region.

\paragraph{Bethe--Heitler limit.}

The perturbative approach to the problem inevitably limits gluon
energies from below. In particular, the gluon spectrum
\eqref{eq:dIdo} describes the Landau-Pomeranchuk-Migdal (LPM)
suppression of the independent Bethe--Heitler radiation and only
applies to gluons with lifetimes larger than the mean free path,
$t>\lambda$, which is equivalent to limiting gluon energies from
above, see \eqref{eq:estt}, 
\begin{equation}
   \omega\> > \> \oBH \sim \mu^2\lambda\>\sim\> \hat{q}\,\lambda^2\,.
\end{equation}
In spite of the fact that the perturbative answer for the quenching
factor \eqref{eq:Fdef} is formally infrared safe, its sensitivity to
the region of small gluon energies may be significant, since the
characteristic energy scale of the problem is strongly reduced by the
bias effect. Indeed, according to \eqref{eq:simple}, to keep the value
of the quenching factor fully under perturbative control we have to
choose the transverse momentum well above
\begin{equation}
\label{eq:BHestim}
   p_\perp \> > \> \pi\,n\> \oBH \>\approx\> 10\>\GeV\,,
\end{equation}
where for the sake of estimate we have taken the vacuum
exponent~\cite{ref3,RHIC} $n\simeq 12\div 13$ and $\oBH\simeq
300\,\MeV$ corresponding to a hot medium with $\hat{q}\simeq
0.2\,\GeV^3$ and $\lambda\simeq \frac14\,\fm$.  At gluon energies
comparable to $\oBH$ Debye screening effect should also become
important thus limiting our ability of doing reliable calculations.

To estimate the absolute minimal value of the quenching factor we take
$p_\perp\la n\,\oBH$ and, keeping in mind that $N(\omega)$ flattens
out at small energies, from \eqref{eq:Fdef} obtain
\[
 \ln\frac{1}{[\,Q(p_\perp)\,]_{\min}} \simeq N(\oBH) 
   \simeq \frac{2\as\, C_R}{\pi}
\sqrt{\frac{2\omega_c}{\oBH}} 
 \simeq  \frac{2\as\,N_c}{\pi}\cdot
\frac{L}{\lambda_R} \>\sim\> \frac{L}{\lambda_R},
\]
where we have used $2\omega_c/\oBH\simeq (L/\lambda)^2$ and introduced
the mean free path for the parton $R$ (quark in our case) which is
related with the gluon mean free path by $C_R\lambda_R= N_c\lambda$.

\section{Illustrations}

\subsection{Shift function and subleading (hard gluon) effects}

If the integrated gluon multiplicity in \eqref{eq:Fdef} depended on a
single variable $\omega/\omega_c$, the quenching factor would have
been a function of a dimensionless ratio
$X=\frac{p_\perp}{n\omega_c}$.  In reality, such a scaling holds only
approximately. Indeed, the gluon radiation spectrum \eqref{eq:dIdo}
and, therefore, the multiplicity $N$, depend on two energy ratios,
$\omega/\omega_c$ and $x=\omega/p_\perp$. The latter takes care, in
particular, of the phase space restriction\footnote{here we use an
approximate equality of $p_\perp$ to the total jet energy} on the
energy of radiated gluons, $dI/d\omega\propto N \propto \Theta(1-x)$.

According to \eqref{eq:xtyp}, the main contribution to quenching
originates from the region $x\sim \omega_1/p_\perp\sim 1/n\ll1$.  The
correction coming from ``hard'' gluons with $x\la 1$ turns out,
however, to be relatively large, $\cO{1/\sqrt{n}}$.
To take into account non-soft corrections, in the numerical evaluation
we supply the spectrum \eqref{eq:dIdo} with a factor
$(1-x)\Theta(1-x)$ as given in~\cite{ref10,ref11}.

The effect of the hard gluon correction is illustrated in
Fig.~\ref{figSS410} where we plot the shift normalized, for
convenience, by $\Delta E=\frac{\pi}{4}\,\alpha\omega_c$ (the BDMPS mean
energy loss~\cite{ref9}).

\begin{figure}[h]
${\displaystyle\frac{S(p_\perp)}{\Delta E}}$
\begin{minipage}{11cm}
 \epsfig{file=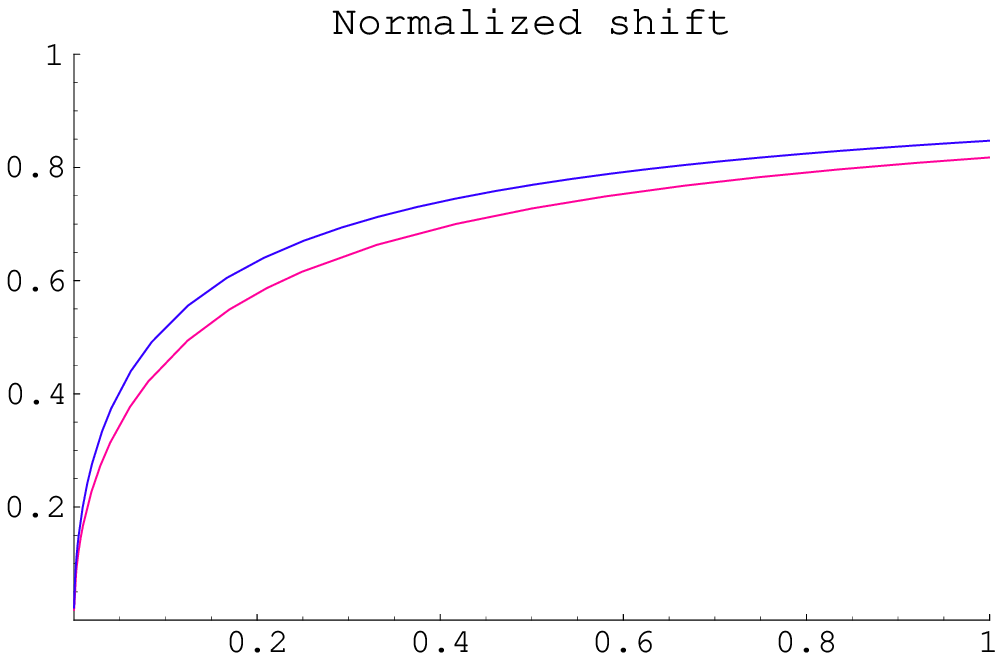, width=10cm}${\quad\displaystyle\frac{p_\perp}{n\,\omega_c}}$
 \end{minipage}
\caption{Shift $S(p_\perp)$ as a function of
$X\equiv\frac{p_\perp}{n\omega_c}$ for $n=4$ (lower curve) and $n=10$
(upper curve).\label{figSS410}}
\end{figure}

These subleading corrections can be treated analytically.  With
account of the phase space restriction, $\omega<p_\perp$, and the
``hard'' factor $(1-x)$ in \eqref{eq:dIdo} the integrated multiplicity
\eqref{eq:Nexpr} gets modified as
\begin{equation}
\label{eq:NexprM}
   N(\omega) \simeq \alpha\left\{  \sqrt\frac{2\omega_c}{p_\perp}\left[\,
\frac1{\sqrt{x}} -2 + \sqrt{x}\,\right]
+\ln2\left[\, \ln x +1-x\,\right]   \right\} ,
 \qquad x=\frac{\omega}{p_\perp}.
\end{equation}
$\left(N(\omega\ge p_\perp)=0\right).$

The energy integral for the shift function \eqref{eq:Sint} then
becomes
\begin{equation}  
\label{eq:Sapprox}
\eqalign{
S(p_\perp) \>&= p_\perp\int_0^1 dx\, N\left(x\,p_\perp\right) \,
e^{- n\,x} \cr
&\approx \alpha\omega_c\cdot \left\{
\sqrt{2\pi\frac{p_\perp}{n\,\omega_c}} \left(1-\frac{2}{\sqrt{\pi\,
n}} + \frac1{2n}\right) -\frac{p_\perp}{n\,\omega_c}\,\ln 2
\left[\,\ln n + \gamma_E -1 + \frac1n \,\right] \right\},
}\end{equation}
where we have omitted exponentially small terms $\cO{\exp(-n)}$.
We see that the hard correction is rather large, $\cO{1/\sqrt{n}}$,
and significantly modifies the behaviour of the shift even in the
strong quenching limit, $X=\frac{p_\perp}{n\omega_c}\ll 1$. 

The exact numerical evaluation of $S(p_\perp)$ is compared with the
approximate formula \eqref{eq:Sapprox} in Fig.~\ref{figSSap}
for $n=4$. 

\begin{figure}[h]
${\displaystyle\frac{S(p_\perp)}{\Delta E}}$ 
\begin{minipage}{11cm}
\epsfig{file=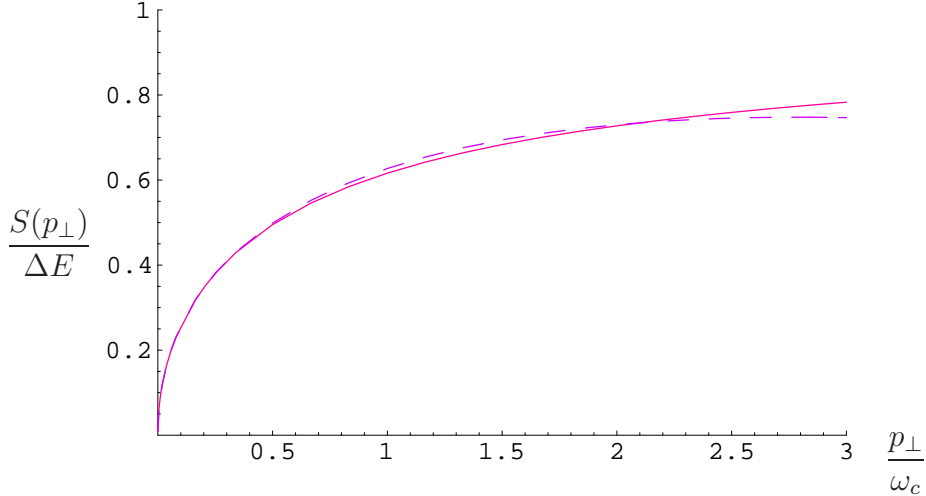, width=10cm}${\quad\displaystyle\frac{p_\perp}{\omega_c}}$
 \end{minipage}
\caption{Shift function for $n=4$ (solid) and its analytic
approximation \eqref{eq:Sapprox} (dashed).\label{figSSap}}
\end{figure}

\noindent
The comparison shows that even for the smallest $n$ value the
approximation based on $p_\perp\ll\omega_c$ turns out to be rather
good up to $p_\perp =2\div 2.5\> \omega_c\sim \sqrt{n}\,\omega_c$.

\subsection{Quenching factors and infrared sensitivity}
The estimate \eqref{eq:BHestim} shows that in the presently available
RHIC range $p_\perp < 6\,\GeV$ a reliable quantitative prediction of
quenching can be hardly made.  It is the soft ``singularity'' of the
LPM spectrum $\omega dI(\omega)/d\omega\propto 1/\sqrt{\omega}$ that
causes instability of the perturbative QCD description.

At the same time, this very same instability makes the study of
quenching in the $p_\perp\sim 20\,\GeV$ range, which is accessible at
RHIC, the more interesting and valuable.  In this region the
characteristic gluon energies are comfortably large, $\omega_1\sim
2\,\GeV$, while the quenching factor at the same time is still quite
sensitive to much smaller energies deep into the infrared domain.

To illustrate this point in Fig.~\ref{figRHIChot} we show the expected
quenching factors for the hot medium ($\hat{q}=0.2\,\GeV^3$, sizes
$L=2$ and $L=5\,\fm$) as a function of an infrared gluon energy
cutoff.  A sharp cutoff is not realistic and can be used only as a
means of quantifying sensitivity of the answer to the low momentum
region. In reality, what matters for quenching is the spectral
properties of the (hot) medium which determine how the gluons with
energies of the order of $\oBH$ are being produced.

\begin{figure}[h]
\begin{center}
\begin{minipage}{7cm}
\epsfig{file=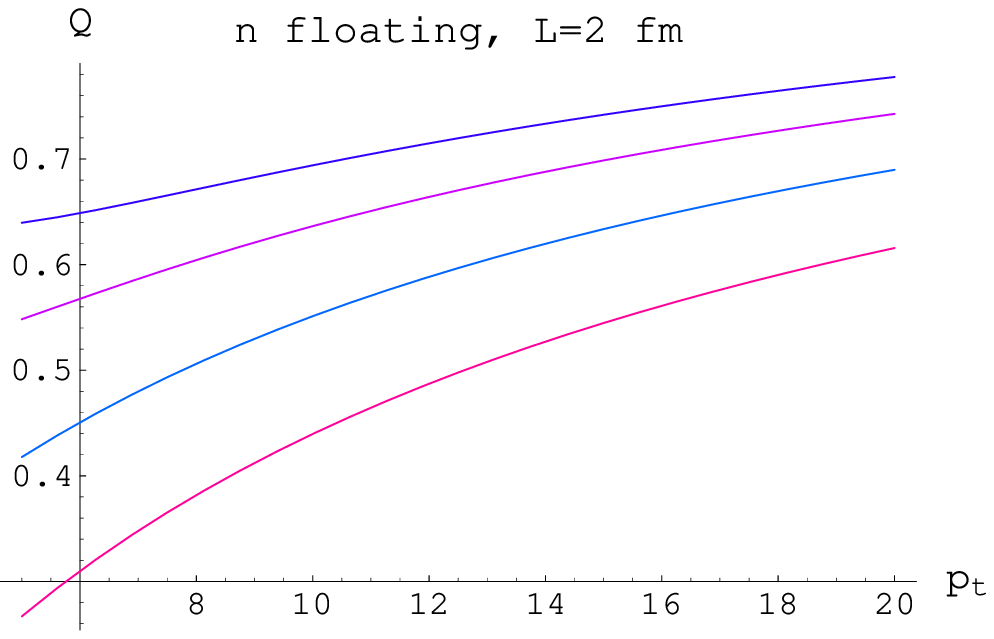, width=7cm, height=7cm}
\end{minipage}
\qquad\qquad
\begin{minipage}{7cm}
\epsfig{file=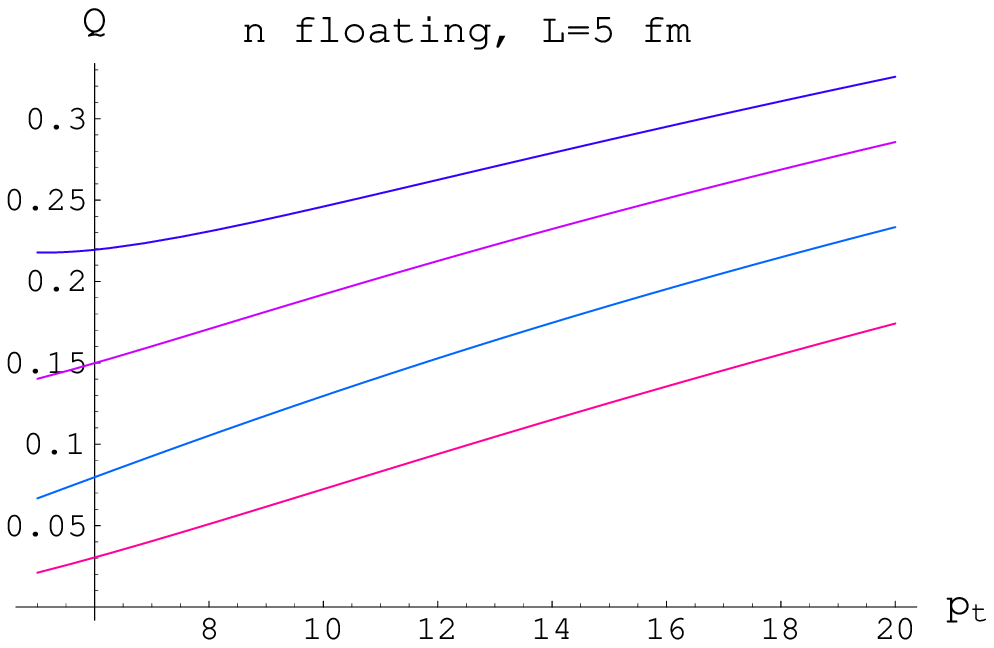, width=7cm, height=7cm}
\end{minipage}
\end{center}
\caption{``Infrared'' dependence of the quenching factor for
hot medium. The curves (from bottom to top) correspond to the gluon
energy cuts 0, 100, 300 and 500~$\MeV$. 
\label{figRHIChot}}
\end{figure}

The label ``$n$ floating'' in Fig.~\ref{figRHIChot} (and below) means
that for this $p_\perp$ range we have used the realistic fit to the
vacuum spectrum provided by the PHENIX collaboration~\cite{ref3},
\[
   \frac{d\sigma^{{\rm vacuum}}(p_\perp)}{dp_\perp^2} \>=\>
\mbox{const}\cdot \left(1.71 \>+\> p_\perp \,[\GeV]\right)^{-12.44} ,
\] 
and evaluated the effective exponent $n(p_\perp)$ according to
\eqref{eq:sigvac}.

In Fig.~\ref{fighotlarge} the quenching factors are shown for large
transverse momenta, where we have set $n=4$, the asymptotic value.  A
mismatch between the $Q$ values at the common point $p_\perp=20\,\GeV$
in Figs.~\ref{figRHIChot} and \ref{fighotlarge} is due to the much
stronger bias effect in the former case:
$n\approx 12 \gg 4$.
  Knowing the $p_\perp$ dependence of the vacuum production cross
  section is, obviously, a necessary ingredient of a reliable
  quenching prediction.

Comparing the two figures we remark that at large transverse momenta
(and smaller $n$) the sensitivity to the infrared physics gets
naturally reduced.

\begin{figure}[h]
\begin{center}
\begin{minipage}{7cm}
\epsfig{file=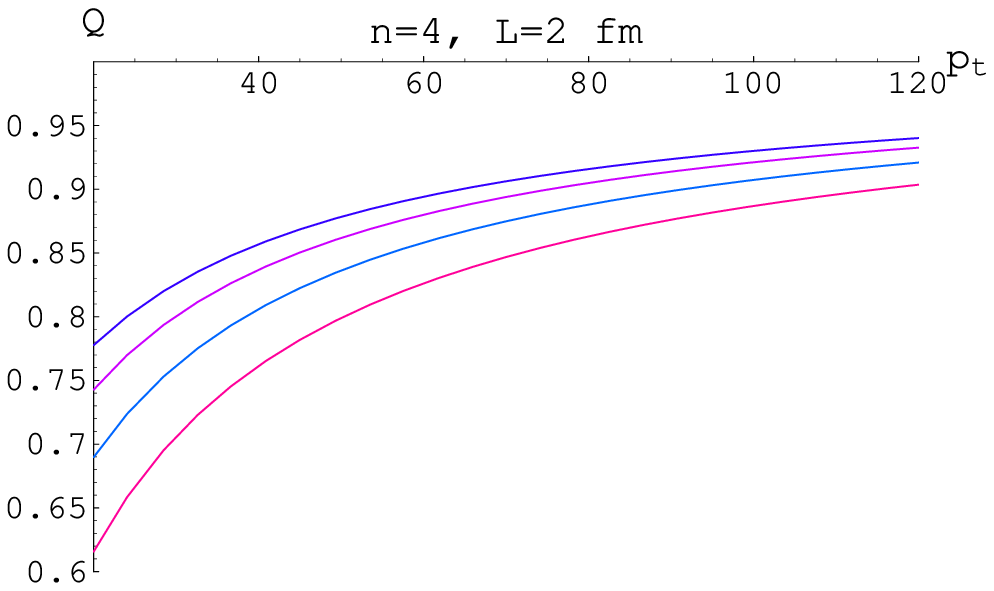, width=7cm, height=7cm}
\end{minipage}
\qquad\qquad
\begin{minipage}{7cm}
\epsfig{file=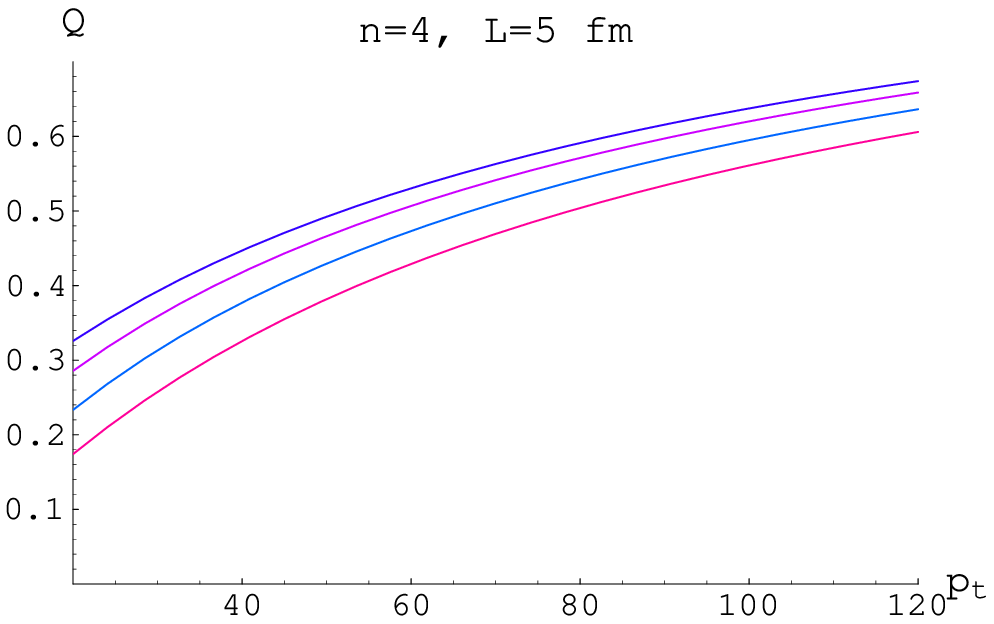, width=7cm, height=7cm}
\end{minipage}
\end{center}
\caption{Quenching factors for hot medium. 
(The curves are the same as in Fig.~\ref{figRHIChot}.)
\label{fighotlarge}}
\end{figure}

In Fig.~\ref{figcold} for the sake of comparison the magnitude
of quenching for a final state matter with the transport coefficient
equal that of the cold nuclear matter, $\hat{q}=\qcold =
0.01\,\GeV^3$, is displayed. 

\begin{figure}[h]
\begin{center}
\begin{minipage}{7cm}
\epsfig{file=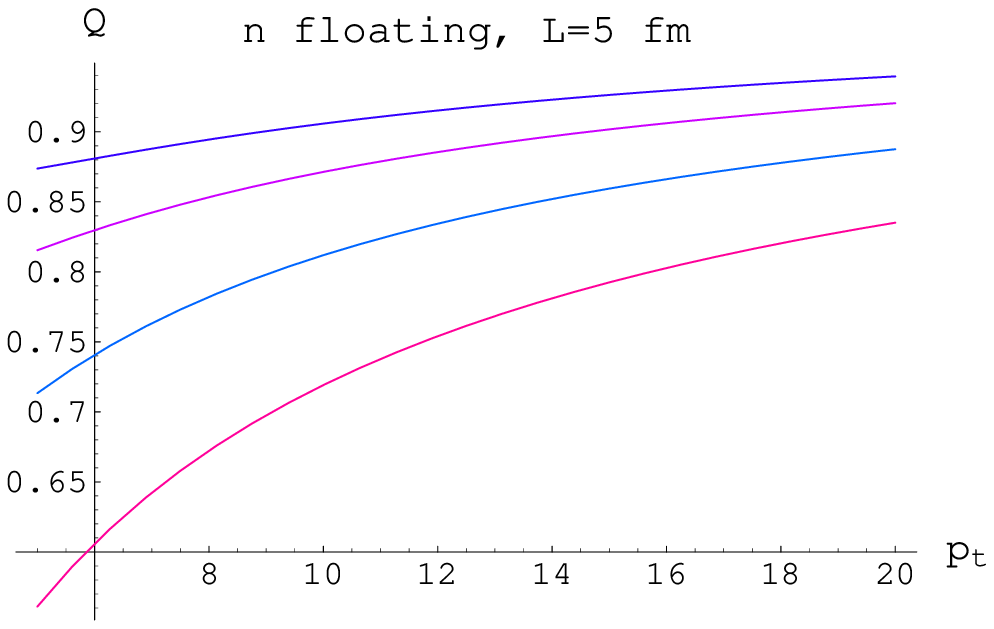, width=7cm, height=7cm}
\end{minipage}
\qquad\qquad
\begin{minipage}{7cm}
\epsfig{file=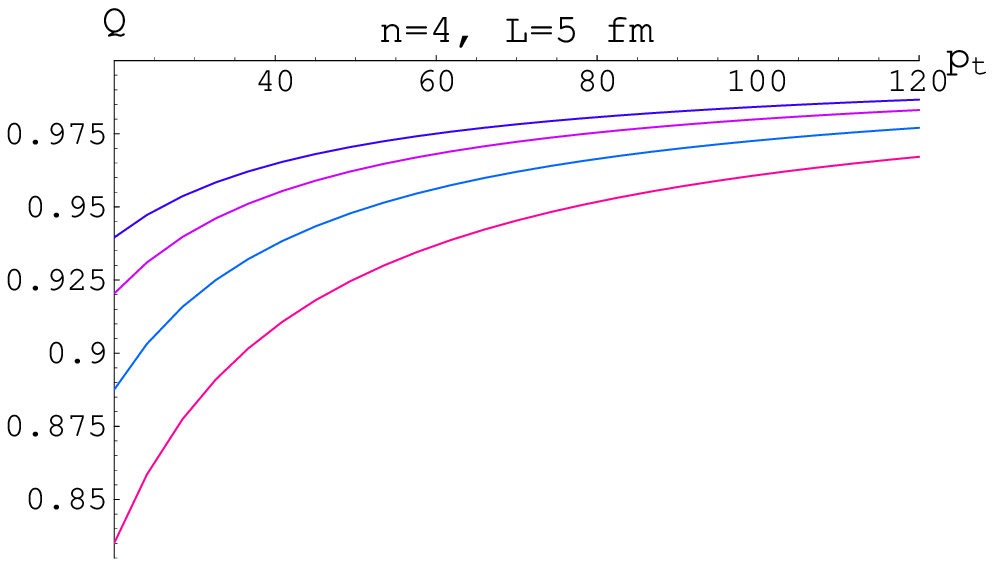, width=7cm, height=7cm}
\end{minipage}
\end{center}
\caption{Quenching factors for cold medium. 
 (The curves are the same as in Fig.~\ref{figRHIChot}.)
\label{figcold}}
\end{figure}

\section{Conclusions}

In this paper we developed the perturbative QCD description of the
suppression (quenching) of inclusive production of hadrons with large
transverse momenta $p_\perp$ in a medium. We did not address the
Cronin effect that {\em enhances}\/ particle yield at moderately large
$p_\perp$.  A simple analysis shows that parton rescattering effects
both in the initial and in the final state (transverse momentum
broadening) should die out as $\cO{\hat{q}L/p_\perp^2}$, to be
compared with the $\cO{L\sqrt{\hat{q}/p_\perp}}$ behaviour
characteristic for quenching. 

We found that quenching is related with the multiplicity of medium
induced (primary) gluons with energies larger than the characteristic
energy $\omega_1$ as
\begin{equation}
-\ln Q(p_\perp) = \int_0^\infty dz\,e^{-z}\> N(\omega_1 z)\,, \qquad
\omega_1(p_\perp)  \>\equiv\> \left[\> \frac{d}{dp_\perp}\,  
\ln \frac{d\sigma^{{\rm vacuum}}}{dp_\perp^2} \>\right]^{-1} 
\>\simeq \frac{p_\perp}{n}\,, \>\> n\gg1\,. 
\end{equation}
One of the results of this study is the observation that the ``shift''
parameter $S$ in the commonly used parametrization
\eqref{eq:shiftdef},
\[
\frac{d\sigma^{{\rm medium}}(p_\perp)}{dp^2_\perp}  
\>\simeq\> \frac{d\sigma^{{\rm vacuum}}(p_\perp+S)}{dp^2_\perp}\,, 
\]
equals neither the {\em mean}\/ medium induced energy loss,
$S=\Delta E \propto L^2$, nor $S=\mbox{const}\cdot L$. In fact, all
over a broad kinematical region where quenching is sizeable, $-\ln
Q(p_\perp) =\cO{1}$, the shift increases with $p_\perp$ as
\[
  S(p_\perp)\>\simeq\> \frac{2\as\,C_F}{\pi}
\sqrt{2\pi\,\omega_c\,\omega_1} \>\propto\> L\,\sqrt{p_\perp}
\]
and equals {\em twice}\/ a typical energy the quark looses to
induced radiation.

We also observed that the particle (jet) quenching phenomenon is
sensitive to the character of the {\em distribution}\/ in the energy
loss $D(\E)$ (in particular, to the energy region
$
  \bar{\E} \sim 
\alpha^2\omega_c \ll \omega_c
$
where $D(\E)$ has a maximum, see \eqref{eq:Dappr}), rather than to its
high-energy tail $\E\sim\omega_c$ which determines the mean energy
loss $\Delta E$.  The latter is dominated by a rare ($\cO{\as}$)
single gluon emission with maximal available energy, $\E\approx \omega
\sim \omega_c$.

Though formally a collinear- and infrared-safe quantity, we found the
quenching factor $Q(p_\perp)$ to be highly sensitive to the region of
small gluon momenta, especially for $p_\perp \la 20\,\GeV$.  The two
effects that contribute to this are the bias ($n\gg1$) and the LPM
suppression which makes the gluon energy spectrum singular at
$\omega\to 0$.
On the one hand, the region where the characteristic gluon energy
$\omega_1$ becomes comparable with $\oBH\sim 300\,\MeV$ (which
regulates the transition from the Landau-Pomeranchuk-Migdal to the
Bethe--Heitler regime) the pure perturbative treatment is hardly
applicable. On the other hand, the experimental (and possibly
theoretical) studies of the momentum region corresponding to
$\omega_1\sim (1\div 2)\,\GeV$
which corresponds to $p_\perp=10\div 20\,\GeV$ 
(for $n\simeq 10$) 
where the Cronin effect is expected to have disappeared,
will provide an important information about the spectral properties of
the final state medium produced in heavy ion collisions.

The final remark concerns our treatment of a medium as uniform and
static, which is obviously not a realistic approximation.  For ``hot''
medium characterized by a large transport coefficient $\hat{q}\sim
0.2\,\GeV^3\simeq 5\,\GeV\,\fm^{-2}$ we estimate the characteristic
lifetime of gluons with energies $\omega_1\sim p_\perp/n$ that
determine quenching as $$
t_1 \sim  \sqrt{\frac{p_\perp}{n\,\hat{q}}}
\quad\sim \left(\>\sqrt\frac{p_\perp}{5\cdot n\> \GeV}\>\right)[\fm] \,.
$$
In the most interesting and practically important situation this
lifetime is small compared with the size of the medium, $t_1\ll L$.
In these circumstances it looks plausible that one will be able to
calculate the quenching factor by employing the basic equations
\eqref{eq:Mells} with the dynamical {\em position-dependent}\/ induced
gluon distribution $dI(\omega,z)/d\omega dz$ given in~\cite{PRC58}.

\end{document}